\documentclass[aps,prd,showkeys,showpacs,amssymb,eqsecnum,
preprintnumbers,nofootinbib,superscriptaddress]{revtex4}

\usepackage{graphicx}
\usepackage{natbib}
\usepackage{latexsym}
\usepackage{amsfonts}

\begin{document}

\newcommand{\vp}{\varphi}
\newcommand{\nn}{\nonumber\\}
\newcommand{\beq}{\begin{equation}}
\newcommand{\eq}{\end{equation}}
\newcommand{\bed}{\begin{displaymath}}
\newcommand{\eed}{\end{displaymath}}
\newcommand{\tC}{\tilde C}
\newcommand{\tB}{\tilde B}
\newcommand{\tA}{\tilde A}
\newcommand{\tc}{\tilde c}
\newcommand{\tb}{\tilde b}
\newcommand{\ta}{\tilde a}
\newcommand*{\cR}{{}^{(D+2)} R}
\newcommand*{\cF}{{}^{(D+2)} F}
\newcommand*{\cA}{{\cal A}}
\newcommand*{\cG}{{}^{(D+2)} G}
\newcommand*{\cX}{{\cal X}}
\newcommand*{\cV}{{\cal V}}
\newcommand*{\D}{{\rm D}}
\newcommand*{\bb}{{\rm b}}
\def\bea{\begin{eqnarray}}
\def\eea{\begin{eqnarray}}

\title{Instability of brane cosmological solutions with flux compactifications }

\author{Masato~Minamitsuji}
\email[Email: ]{Masato.Minamitsuji"at"physik.uni-muenchen.de}
\affiliation{Arnold-Sommerfeld-Center for Theoretical Physics, Department f\"{u}r Physik, Ludwig-Maximilians-Universit\"{a}t, Theresienstr. 37, D-80333, Munich, Germany}

\begin{abstract}
We discuss the stability of the higher-dimensional de Sitter (dS) brane solutions
with two-dimensional internal space in the Einstein-Maxwel theory.
We show that an instability appears in the scalar-type perturbations
with respect to the dS spacetime.
We derive a differential relation which has the very similar structure
to the ordinary laws of thermodynamics as an extension of the work
for the six-dimensional model \cite{ksm}. 
In this relation, the area of dS horizon (integrated over the two internal dimensions)
exactly behaves as the thermodynamical entropy.
The dynamically unstable solutions are in the thermodynamically unstable
branch.
An unstable dS compactification either 
evolves toward a stable configuration or
two-dimensional internal space is decompactified.
These dS brane solutions are equivalent to
the accelerating cosmological solutions in the six-dimensional
Einstein-Maxwell-dilaton theory via dimensional reduction.
Thus, if the seed higher-dimensional solution is unstable,
the corresponding six-dimensional solution is also unstable.
From the effective four-dimensional point of view,
a cosmological evolution from an unstable cosmological solution in
higher dimensions may be seen as a process of the transition
from the initial cosmological inflation to the current dark
energy dominated Universe.
\end{abstract}

\preprint{LMU-ASC 05/08}
\maketitle

\section{Introduction}

Six-dimensional braneworld models have attracted particular interests 
in recent years\cite{p}.
From the cosmological aspects, these models may be useful
as a way of resolution of the cosmological constant problem,
since a codimension two brane helps the sideslip of the brane
vacuum energy into the bulk \cite{cc+}.
It has been pointed out that the original proposal of
the resolution of this problem actually does not work well \cite{cc-}.
Nevertheless, six dimensional models has been recognized as an important
playground to unsderstand cosmology and gravity in higher-dimensional
theory with non-trivial fluxes.
The flux stabilization of extra dimensions would be a powerful tool
to obtain realistic phenomenology and cosmology from string theory.
In the simplest realization of the flux compactifications in six dimensions,
the internal space has the shape of a rugby ball \cite{cc+,cc-}, 
where codimension two branes are located at the positions of the poles.
The warped generalizations of the rugby ball solutions have also
been reported in the context of the six-dimensional Nishino-Sezgin (Salam-Sezgin),
gauged supergravity
\cite{ggp} (see \cite{ns} for the original supergravity)
and pure Einstein-Maxwell theory \cite{msyk}.
\footnote{In Ref. \cite{cdgv}, a study on warped codimension-two 
braneworld solutions was presented 
on the analogy of the classical mechanics.}

It also has been recognized that
a 3-brane in six or higher dimensions 
generically have the problems on localication of matter
on the brane due to its stronger self-gravity.
One well motivated way to circumvent this problem
is to regularize the brane, by taking the microscopic
structure of the brane into account.
Several ways of regularization of codimension two branes
have been proposed in \cite{thick,pst,vc}.
Based on these regularizations,
low energy cosmology \cite{vc,cosmo,cck} and effective
gravity on the brane \cite{grad} have been studied.
On the other hand, the exact solutions \cite{6dt, km, shock,btz} will help
to obtain unique observational/experimental predictions from six dimensions.

Stability of six-dimensional flux compactifications
is an important issue and
several analyses of linear perturbations have been reported.
It has been reported that the Minkowski brane solutions
in the supergravity \cite{ns} are marginally stable \cite{st1}
and those in the Einstein-Maxwell theory are stable \cite{st2}. 
On the other hand, in the de Sitter (dS) brane solutions in the
Einstein-Maxwell theory an instability 
appears in the scalar sector of perturbations
with respect to the symmetry of dS spacetime
for a relatively higher brane expansion rate \cite{ksm}.
This type of instability is commonly known in the
dS spacetime~\cite{kp}
with an internal space compactified by a flux~\cite {fr} 
\footnote{Other instabilities in the quadrupole or higher multipole modes
were reported, which gives rise to deformation of
the internal space geometry, see e.g., Ref. \cite{martin}.
However, these instabities appear in dS compactifications
with more than four-dimensional internal space
and are not relevant for the models discussed in this paper.}
We will see that such an instability also appears
in the dS brane solutions in higher-dimensional Einstein-Maxwell
theory. 
The important fact is that
a class of the six-dimensional Einstein-Maxwell-dilaton theories
has the equivalent structure to that of the higher-dimensional
Einstein-Maxwell theory via dimenional reduction \cite{km}.
Thus instability of a solution in the higher dimensional theory 
suggests that of the corresponding solution in six-dimensional
theory.

Also in Ref. \cite{ksm}, an important relation which has
very similar structure to the ordinary laws of thermodynamics was found
in the dS brane solutions in the six-dimensional Einstein-Maxwell theory.
In this relation, the area of dS horizon (integrated over the internal space)
exactly behaves as the usual thermodynamical entropy.
It was shown that
dynamically unstable solutions are also
{\it thermodynamically} unstable.
This may be seen as an example to support
the conjecture that claims the equivalence of these two
instabilities \cite{gm}, which originally has been discussed
for black brane solutions.
We will see that such a thermodynamical relation
can be easily extended to the case the higher
dimensional dS brane solututions.

The dynamical evolution from unstable dS flux compactifications \cite{fr}
have been investigated in e.g., Ref. \cite{krish}.
Based on these arguments, we will discuss the fate of unstable
dS brane solutions and corresponding cosmological
solutions in six dimensions.
We will see that an initially unstable dS brane solutions evolves to
another stable dS or anti-de Sitter (AdS) brane solution
unless the internal space is decompactified.

This article is organized as follows.
In the the section II, we review the higher-dimensional dS brane solutions
and cosmological solutions in the general six-dimensional Einstein-Maxwell-dilaton
theory
via dimensional reduction from the dS brane solutions. 
In the section III, we analyze the stability and cosmological evolutions
of the higher-dimensional dS compactifications.
In the section IV, we discuss the possible fate of unstable solutions
in the higher dimensional theory from the six-dimensional perspectives.
In the section V, we shall close this article.



\section{de Sitter brane solutions with flux compactification}

We start with the $(D+2)$-dimensional Einstein-Maxwell action
\begin{eqnarray}
S^{(D+2)}=\int d^{D+2} X \sqrt{-G}
\Big[\frac{1}{2}\left(
{}^{(D+2)} R-2\Lambda\right)
-\frac{1}{4}\cF_{MN}\cF^{MN}
\Big],
\label{6+n-d-action}
\end{eqnarray}
and the $D$-dimensional brane actions contain
the tension $\sigma_{i}$ where $(i=1, 2)$.
We set the $(D+2)$-dimensional gravitational scale
$M_{D+2}=1$, unless it should be shown explicitly. 
There are
the warped de Sitter (dS) brane solutions
with two-dimensional internal space compactified by
a magnetic flux.
The bulk metric is given by
\begin{eqnarray}
ds^2_{(D+2)}=
\xi^{(2-2\gamma^2)/(1+\gamma^2)}
\gamma_{\alpha\beta}dz^{\alpha}dz^{\beta}
+\frac{\xi^{-2\gamma^2/(1+\gamma^2)}}{2\Lambda}
\Big[
\frac{d\xi^2}{h(\xi)}+\beta^2h(\xi)d\theta^2
\Big],
\label{intermsofxi}
\end{eqnarray}
where
\begin{eqnarray}
h(\xi)&:=&\frac{(1+\gamma^2)^2}{4(5-\gamma^2)}
\Big[
-\xi^{2/(1+\gamma^2)}
+\frac{1-\alpha^{8/(1+\gamma^2)}}
      {1-\alpha^{(3+\gamma^2)/(1+\gamma^2)}}
\frac{1}{\xi^{(3-\gamma^2)/(1+\gamma^2)}}-
\frac{\alpha^{(3+\gamma^2)/(1+\gamma^2)}
      (1-\alpha^{(5-\gamma^2)/(1+\gamma^2)})}
     {1-\alpha^{(3+\gamma^2)/(1+\gamma^2)}}
\frac{1}{\xi^{6/(1+\gamma^2)}}
\Big]
\nonumber\\&&
+\frac{\lambda}{\Lambda}
\frac{1+\gamma^2}{6+2\gamma^2}
\xi^{2\gamma^2/(1+\gamma^2)}
\Big[ 1-\frac{1}{\xi^{(3+\gamma^2)/(1+\gamma^2)}} \Big]
\Big[ 1-\frac{\alpha^{(3+\gamma^2)/(1+\gamma^2)}}
{\xi^{(3+\gamma^2)/(1+\gamma^2)}} \Big],
\end{eqnarray}
has two positive roots at $\xi=1,\alpha$.
The $D$-dimensional metric
$\gamma_{\alpha\beta}$ is that of dS spacetime,
which satisfies
$R_{\alpha\beta}[\gamma]=(2/(D-2))\lambda\gamma_{\alpha\beta}$
(The expansion rate with respect to the dS proper time is given by
$H^2= 2\lambda/(D-1)(D-2)$).
In the above expression, we have also introduced
\begin{eqnarray}
\gamma:=\sqrt{\frac{D-4}{D}}.\label{def_gamma}
\end{eqnarray}
As we will see later, $\gamma$ is being the dilatonic coupling
in a class of the Einstein-Maxwell-dilaton theories which
are equivalent to the higher-dimensional Einstein-Maxwell
theory via dimensional reduction (see Eq. ({\ref{6d_action})).
After a dimensional reduction, $D$ needs not to be restricted 
to be an integer.
We restrict $\alpha \leq \xi\leq 1$ and assume that 
$\theta$ has the period $2\pi$.
The field strength is now given by
\begin{eqnarray}
\cF_{\xi\theta}&=&\frac{
\beta}{\sqrt{2\Lambda}}
\frac{Q}{\xi^{2(2+\gamma^2)/(1+\gamma^2)}}
\label{fs}
\end{eqnarray}
where the magnetic charge $Q$ is
\begin{eqnarray}
Q&:=&
\alpha^{\frac{3+\gamma^2}{2(1+\gamma^2)}}
\Big(
 \frac{3+\gamma^2}{5-\gamma^2}
 \frac{1- \alpha^{(5-\gamma^2)/(1+\gamma^2)}}
      {1- \alpha^{(3+\gamma^2)/(1+\gamma^2)}}
-\frac{2}{1+\gamma^2}\frac{\lambda}{\Lambda}
 \Big)^{1/2}.\label{mag}
\end{eqnarray}
$\lambda$ is bounded from above:
\begin{eqnarray}
\lambda
\leq
\lambda_{\text{max}}(\alpha)
&:=&
\frac{(1+\gamma^2)(3+\gamma^2)}
     {2(5-\gamma^2)}
\frac{1- \alpha^{(5-\gamma^2)/(1+\gamma^2)}}
     {1- \alpha^{(3+\gamma^2)/(1+\gamma^2)}}
\Lambda.
\label{upperboundlambda}
\end{eqnarray}
The constant $\beta$ controls deficit angles
at $\xi=1$ and $\xi=\alpha$, which are given, respectively, by
$\delta_i
=2\pi\big[1-\beta
\big|h'(\xi)\big|_{\xi=\xi_i}/2
\big]$, where $i=+,-$ represents the branes
$\xi_+=1$ or $\xi_-=\alpha$, respectively.
The conical deficit corresponds to a codimension two brane and the tension is given by
$\sigma_i/M^{D}_{D+2} = \delta_i$.

\section{Instability of dS brane solutions}

\subsection{Dynamical instability}

We briefly discuss the stability of the higher-dimensional 
dS brane solutions against the scalar-type
perturbations with respect to the symmetry of $D$-dimensional dS spacetime.
It is instructive to focus on the case $\alpha =1$, where
analytic solutions for perturbations are available.
Introducing a new bulk coordinate
$\xi=((1-\alpha)\sin w +(1+\alpha))/2$,
in order to resolve the spacetime structure in the limit $\alpha=1$,
the $(D+2)$-dimensional metric Eq. (\ref{intermsofxi})
in the case $\alpha= 1$
can be expressed as
\begin{eqnarray}
ds_{(D+2)}^2
&=&\gamma_{\mu\nu}dx^{\mu}dx^{\nu}
+\frac{1+\gamma^2}
     {2\Lambda(1+\gamma^2)-(3+\gamma^2)\lambda}
\Big(dw^2
+\tilde \beta^2\cos^2 w d\theta^2\Big),
\end{eqnarray}
where $\tilde \beta$ is a constant related to $\beta$.
The vector field is also rewritten into the form of
$A_{\theta}\propto \sin w$. 

We discuss scalar perturbations and work in the longitudinal gauge.
\begin{eqnarray}
ds_{(D+2)}^2
&=&
(1+2 \Omega_2(w,x^{\mu}))\gamma_{\mu\nu}dx^{\mu}dx^{\nu}
+\frac{1+\gamma^2}
     {2\Lambda(1+\gamma^2)-(3+\gamma^2)\lambda}
\nonumber\\
&\times&
\Big[\big(1+2(\Omega_1(w,x^{\mu})
           +\Omega_2(w,x^{\mu}))\big) dw^2
+\big(1+2(-\Omega_1(w,x^{\mu})
-\frac{3+\gamma^2}{1-\gamma^2}\Omega_2(w,x^{\mu}))\big)
\tilde\beta^2\cos^2 w d\theta^2
\Big].
\nonumber\\
&&
\end{eqnarray}
Note that we assume that the all the perturbation mode 
are axial symmetric and drop the dependence on
the angular coordinate $\theta$.
Then, the magnetic field perturbation is written as
\begin{eqnarray}
a_{\theta}^{(1)}
=
\frac{2\big((1+\gamma^2)\Lambda-2\lambda\big)}
     {2(1+\gamma^2)\Lambda-(3+\gamma^2)\lambda}
 \Big(
-\big(2\Omega_1+\frac{4}{1-\gamma^2}\Omega_2\big)
 +\frac{\Omega_{1,w}}{\tan w}
 \Big).
\end{eqnarray}
We expand in terms of eigenmodes
$\Omega_i =\sum_n \chi_n(x^{\mu}) \omega_{i,n}(w) (i=1,2)$,
where $\Box_{D}\chi_n (x^{\mu})=\mu_n^2 \chi_n^{\mu}$, where
$\mu_n$ represents the effective $D$-dimensional mass parameter
of the $n$-th Kaluza-Klein (KK) mode
and $\Box_{D}$ is d'Alembertian 
with respect to $D$-dimensional dS metric $\gamma_{\mu\nu}$.
These two variables obey a couple of equations of motion.
The solutions for the bulk mode $\omega_i$ are given
in terms of the Legendre functions of order $\nu_{\pm}$:
$\nu_{\pm} (\nu_{\pm} + 1)=\lambda_{\pm}$, where
\begin{eqnarray}
\lambda_{\pm}
&=&1
+\frac{(1+\gamma^2)\mu^2}{2(1+\gamma^2)\Lambda-(3+\gamma^2)\lambda}
\pm
\sqrt{1+\frac{2(1+\gamma^2)(3+\gamma^2)
             \big((1+\gamma^2)\Lambda-2\lambda\big)\mu^2}
{\big(2(1+\gamma^2)\Lambda-(3+\gamma^2)\lambda\big)^2}}.
\end{eqnarray}
We impose the regularity at both the boundary branes and then
we obtain the eigenmodes from the conditions $\nu_{\pm}=n(=0,1,2)$.
For the $(+)$-branch, the mass of the lowest mode is given by
\begin{eqnarray}
 \mu_0^2
= \big(1+\gamma^2\big)
\Big(1-\frac{(3+\gamma^2)\lambda}
            {(1+\gamma^2)^2\Lambda}
\Big)\Lambda
\,.
\end{eqnarray}
Clearly, for
\begin{eqnarray}
\frac{\lambda}{\Lambda}
>\frac{\lambda_{\rm ins}}{\Lambda}
:=
\frac{(1+\gamma^2)^2}
     {3+\gamma^2},\label{yoko}
\end{eqnarray}
the lowest mode becomes tachyonic.
Note that the upper bound on the brane expansion rate is
now given from Eq. (\ref{upperboundlambda})
\begin{eqnarray}
\frac{\lambda_{\rm max}}{\Lambda}
=
\frac{1+\gamma^2}
     {2}.\label{max}
\end{eqnarray}
The lower tachyonic mass is bounded from the below
\begin{eqnarray}
\mu_0^2
\geq 
-\frac{(1-\gamma^2)\Lambda}{2}\,,
\end{eqnarray}
and hence this mode disappears in the limit $\gamma \to 1$ ($D\to \infty$),
which corresponds to the Nishino-Sezgin (Salam-Sezgin) gauged supergravity theory in the equivalent 
six-dimensional picture (See Sec. IV).


Before closing this subsection, we shall comment on the stability against 
tensor- and vector-type
perturbations with respect to the symmetry of the dS spacetime.
Stability of the higher-dimensional dS brane solutions
against the tensor perturbations
has been shown in Ref. \cite{km}, irrespectively of $\alpha$ and $\lambda$.
Note that stability of the six-dimensional solutions against the tensor perturbations
was originally confirmed in Ref. \cite{ksm}.
In Ref. \cite{ksm}, stability against the vector perturbations was also shown.
Both the stability against the tensor perturbations and 
the appearance of an instability against the scalar perturbations
are the features that are not relevant for the number of dimensions
of dS spacetime.
Thus, it is quite natural to expect that the dS brane solutions
are stable against the vector perturbations,
irrespectively of the number of dimensions.


\subsection{dS thermodynamics}

In Ref. \cite{ksm}, for the dS brane solutions in the six-dimensional Einstein-Maxwell
theory,
a differential relation which has the very similar
structure to the ordinary laws of thermodynamics
has been derived.
In this relation, the area of the dS horizon
(cosmological horizon) integrated over the internal space
behaves like the thermodynamical entropy.
Then, it was shown that the dynamically unstable solutions
were also {\it thermodynamically unstable}, namely 
these two instabilities were equivalent in such a system.
As we will see in this subsection,
the {\it  dS thermodynamics} can be extended to the cases of
higher dimensional dS brane solutions.
The area of dS horizons
(divided by the area of $(D-2)$-sphere $\Omega_{D-2}$)
is given by
\begin{eqnarray}
{\cal A}
=
\frac{\beta H^{-(D-2)}}{(2\Lambda)}
 \frac{\pi(D-2)}{D-1}
\Big(
1-\alpha^{2(D-1)/(D-2)}
\Big).
\end{eqnarray} 
We also find conserved quantities,
the magnetic flux
\begin{eqnarray}
\phi:=\sqrt{2\Lambda}\int_{\alpha}^1 d\xi\int_0^{2\pi}d\theta
{}^{(D+2)}F_{\xi\theta}
=\frac{\pi\beta(D-2)}{(D-1)}
\frac{Q}{\alpha^{2(D-1)/(D-2)}}
\Big(
1-\alpha^{2(D-1)/(D-2)}
\Big). \label{flux2k}
\end{eqnarray}
and
\begin{eqnarray}
&&\beta_+
=\frac{-h_{,\xi}(\xi=1)}{2}\beta,\quad
\beta_-
=\frac{h_{,\xi}(\xi=\alpha)}{2}\beta\,,
\label{betapm}
\end{eqnarray}
which are directly related to the brane tensions,
located at $\xi=1$ and $\xi=\alpha$, as
$\sigma_{\pm}=2\pi(1-\beta_{\pm})$.
It is straightforward to find the relation
\begin{eqnarray}
\beta_+ +\alpha^{2D/(D-2)} \beta_-
=\frac{1}{2\pi}
\Big(
 Q\phi
+(D-1)H^D {\cal A}
\Big).\label{id}
\end{eqnarray}
Note that $H^2=(2/(D-1)(D-2))\lambda$ is the expansion rate of
the dS spacetime, with respect to the dS proper time.

The area of dS horizon is related to the area of dS horizon
\begin{eqnarray}
S_E&=&
-\int d^{D+2}X \sqrt{G}
\Big(
\frac{1}{2}{}^{(D+2)}R
-\Lambda
-\frac{1}{4}{}^{(D+2)}F_{AB}
{}^{(D+2)}F^{AB}
\Big)
=-(D-1)\Omega_D {\cal A}
\end{eqnarray}
where $\Omega_D$ is the area of $D$-sphere.
The stationary solutions correspond to 
the points where the Euclidean action has
the maximum with respect to the variable $(\alpha,\lambda)$.
The observer at the $(+)$-brane cannot adjust the tension
of the $(-)$-brane.
It is useful to define the intensive quantities as
$\tilde {\cal A} :={\cal A}/\beta_-$,
$\tilde \phi:=\phi/\beta_-$ and $\eta:=\beta_+/\beta_-$.
The similar quantities divided by $\beta_+$
can be defined for the observer on $(-)$-brane.
Of course, these two points of views are equivalent.
In the later discussion, we will take the point of view from the $(+)$-brane.
An extremal condition $\big(\partial S_E/\partial \alpha\big)_{H}=0$
gives
\begin{eqnarray}
&&\tilde \phi
 =2\pi\Big(
     \frac{2D}{D-1}
     \Big)
\frac{\alpha^{(D+2)/(D-2)}}
    {\big(\frac{\partial Q}{\partial \alpha}\big)_H}
\,.\label{sta1}
\end{eqnarray}
The other extremal condition
$\big(\partial S_E/\partial H\big)_{\alpha}=0$ gives
\begin{eqnarray}
\Big(H
\big(\frac{\partial Q}{\partial H}\big)_{\alpha}
-D Q\Big)\phi
=-2\pi D
\Big(
\beta_+ 
+
\beta_-\alpha^{2D/(D-2)}
\Big)\,.\label{sta2}
\end{eqnarray}
These two conditions determine the stationary points
$(\alpha_e,H_e)$,
which are analytically continued to the stationary solutions
in the original Lorentzian theory.
They can be reduced to the following relations
\begin{eqnarray}
&&(D-1) H_{e}^D {\cal A}
=2\pi \Big(\beta_+
+\beta_-\alpha_e^{2D/(D-2)}
\Big)
-\phi Q_e
\nonumber \\
&&\phi d Q_e=
2\pi \beta_-
d\alpha_e^{2D/(D-2)}
-(D-1) {\cal A}dH^D_e.
\end{eqnarray}
The first one is the repetition of Eq. (\ref{id}).
From these relations, 
in terms of the {\it intensive} variables,
the following differential relation is obtained
\begin{eqnarray}
d\big(-{\tilde {\cal A}}\big)
=\frac{1}{(D-1)H^{D}}
\Big(
2\pi
d(-\eta)
+Q d\tilde \phi
\Big).
\label{1st}
\end{eqnarray}
This relation has
the same form as the first law of thermodynamics:
For fixed $\beta_-$, we obtain
\begin{eqnarray}
 d(-\eta)=\frac{1}{2\pi\beta_-}d\sigma_+.
\end{eqnarray}
We can see that the change of $(-\eta)$ corresponds to that of
the {\it internal energy} of the system.
The magnetic flux $\tilde \phi$ and magnetic charge $Q$ can be
analogies of the {\it volume} and {\it pressure}, respectively. 
The dS expansion rate $H^D$ can be seen as the
thermodynamical {\it temperature}.
So, $(-\tilde {\cal A})$ corresponds to the thermodynamical 
{\it entropy}.

For a fixed value of conserved quantity $(\tilde \phi,\eta)$,
the area of a dS horizon is double-valued:
there are two possible branches, thermodynamicall unstable (low entropy)
and stable (high entropy) branches.
We can see that the dynamically unstable solutions are
belonging to the low entropy branch.
To show this,
we consider the thermodynamical stability condition
$\delta^2 (-{\cal A})<0$
:
This requires the inequalities
\begin{eqnarray}
&&\Big(\frac{\partial\eta}{\partial H^D}
\Big)_{Q}
<\Big(\frac{\partial\eta}{\partial H^D}
\Big)_{\tilde\phi}<0,
\quad
\Big(\frac{\partial\tilde\phi}{\partial\big(Q H^{-D}\big)}
\Big)_{H}
<
\Big(\frac{\partial\tilde\phi}{\partial\big(Q H^{-D}\big)}
\Big)_{\eta}<0.
\end{eqnarray}
For instance,
the first condition states that the {\it specific heats} are positive.
We can explicitly obtain
\begin{eqnarray}
&&\Big(\frac{\partial\eta}{\partial H^D}
\Big)_{\tilde\phi}
=
\frac{1}{\Big(\frac{\partial\tilde \phi}
            {\partial\alpha}\Big)_{H}}
\frac{\partial (\eta,\tilde \phi)}
     {\partial (H^D, \alpha)},
\nonumber\\
&&\Big(\frac{\partial\big(Q H^{-D}\big)}
            {\partial\tilde\phi}
\Big)_{\eta}
=-\frac{1}{H^{2D}}
\Big(\frac{\partial (\eta,\tilde \phi)}
     {\partial (H^D, \alpha)}\Big)^{-1}
\Big[\Big(H^D\Big(\frac{\partial Q}{\partial H^D}\Big)_{\alpha}-Q\Big)
    \Big( \frac{\partial \eta}{\partial \alpha}\Big)_{H}
  -H^D \Big(\frac{\partial Q}{\partial \alpha}\Big)_{H}
\Big( \frac{\partial \eta}{\partial H^D}\Big)_{\alpha}
\Big].
\end{eqnarray}
It is straightforward to see that
those quantities are negative for $\lambda<\lambda_{\rm crit}(\alpha)$ for each
$\alpha$. 
Here $\lambda_{\rm crit}(\alpha)$ represents the position
of the critical curve, on which 
a map from $(\alpha,H)$ plane to $(\eta,\tilde\phi)$ plane breaks down:
\begin{eqnarray}
\frac{\partial (\eta,\tilde \phi)}
     {\partial (\alpha,H)}
=0\,.
\end{eqnarray}
The critical curve can be analytically obtained (in terms
of $\lambda$) as
\begin{eqnarray}
\frac{\lambda_{\rm crit}}{\lambda_{\rm max}}
&=&\frac{1}{4
     \big(3+\gamma^2\big)
   \big(1-\alpha^{(5-\gamma^2)/(1+\gamma^2)}\big)
   \big(1-\alpha^{(3+\gamma^2)/(1+\gamma^2)}\big)^2
}
\nonumber\\
&\times&
\Big\{
 11
+4\alpha^{-1+\frac{6}{1+\gamma^2}}
-25\alpha^{1+\frac{2}{1+\gamma^2}}
-4\alpha^{2+\frac{4}{1+\gamma^2}}
+25\alpha^{\frac{8}{1+\gamma^2}}
-11\alpha^{1+\frac{10}{1+\gamma^2}}
+6 \gamma^2
-4\alpha^{-1+\frac{6}{1+\gamma^2}}\gamma^2
\nonumber\\
&-&22\alpha^{1+\frac{2}{1+\gamma^2}}\gamma^2
+4\alpha^{2+\frac{4}{1+\gamma^2}}\gamma^2
+22 \alpha^{\frac{8}{1+\gamma^2}}\gamma^2
-6\alpha^{1+\frac{10}{1+\gamma^2}}\gamma^2
-\gamma^4
-\alpha^{1+\frac{2}{1+\gamma^2}} \gamma^4
+\alpha^{\frac{8}{1+\gamma^2}}\gamma^4
+\alpha^{1+\frac{10}{1+\gamma^2}}\gamma^4
\nonumber\\
&-&
\Big(
 1
+4\alpha^{-1+\frac{6}{1+\gamma^2}}
-4\alpha^{1+\frac{2}{1+\gamma^2}}
-\gamma^2
-\alpha^{\frac{8}{1+\gamma^2}}
(1-\gamma^2)\Big)
\sqrt{(1-\gamma^2)^2(1+\alpha^{2+\frac{4}{1+\gamma^2}})
+2\alpha^{1+\frac{2}{(1+\gamma^2)}}
(17+14\gamma^2+\gamma^4 )}
\Big\}\,,
\nonumber\\
&&
\end{eqnarray}
where the maximum value of the flux $\lambda_{\rm max}$
is given in Eq. (\ref{max}).
In the case that $\gamma\to 0$ (the six-dimensional limit),
we recover the result in \cite{ksm}
\begin{eqnarray}
\frac{\lambda_{\rm crit}}{\lambda_{\rm max}}
&=&\frac{1}{12(1+\alpha+\alpha)^2(1+\alpha+\alpha^2+\alpha^3+\alpha^4+\alpha^5)}
\Big[
 11+33\alpha+\ 66\alpha^2+85\alpha^3+90\alpha^4
+85\alpha^5+66\alpha^6+33\alpha^7+11\alpha^8
\nonumber\\
&-&(1+\alpha)\big(1+2\alpha+4\alpha^2+2\alpha^3+\alpha^4\big)
\sqrt{1+34\alpha^3+\alpha^6}
\Big]\,.
\end{eqnarray} 
In Ref. \cite{ksm}, it has been seen that
for the solutions belonging to the above critical curve
the lowest mode of the scalar perturbations
exactly becomes massless and thus this curve
gives the border between the families of the stable
and unstable solutions.
This must be true for the higher dimensional dS solutions.
For the general $\gamma$ (hence $D$),
the limit $\alpha\to 1$ gives
\begin{eqnarray}
\frac{\lambda_{\rm crit}}{\lambda_{\rm max}}
=\frac{2(1+\gamma^2)}{3+\gamma^2}\,.
\end{eqnarray}
namely $\lambda_{\rm crit}=\lambda_{\rm inst}$, which
is defined in Eq. (\ref{yoko}).

\subsection{Cosmological evolutions}

In this subsection, we see the cosmological evolutions from
an unstable dS solution.
We assume that the $D$-dimensional geometry keeps
homogeneity and isotropy after the deviation
from the exact dS geometry.
In our perturbation analysis in subsection III.A,
we assume that $\Omega_1$ and $\Omega_2$ are functions
of the space and time in $D$-dimensional dS space time
(More precisely, we expanded them in terms of scalar 
harmonic functions defined on the $D$-dimensional dS spacetime).
However, in order to discuss the cosmological evolutions,
we assume the perturbations as the functions
of only the time, in order to keep the homogeneity and
isotropy of $D$-dimensional spacetime.
This assumption will also be essential to discuss the 
cosmology in terms of the equivalent six-dimensional theory in Sec. IV.

We firstly focus on the case $\alpha=1$.
In this case, it is natural that the bulk shape is not deformed during cosmological evolution
(and hence always $\alpha=1$).
We consider the homogeneous evolution of the external and internal
space
\begin{eqnarray}
ds^2
=
-dt^2+a(t)^2 d{\bf x}_{D-1}^2
+
c(t)^2
\Big[ dw^2+\tilde\beta^2\cos^2 w d\theta^2
\Big].
\end{eqnarray}
For the conserved magnetic flux,
\begin{eqnarray}
\phi=\sqrt{2\Lambda}\int dwd\theta {}^{(D+2)} F_{w\theta}
=4\pi\tilde \beta
 \frac{\sqrt{1-\frac{\lambda}{\Lambda}\frac{D}{D-2}}}
      {1-\frac{\lambda}{\Lambda}\frac{D-1}{D-2}}
\label{flux2}
\end{eqnarray}
the radion equation of motion is typically given by
\begin{eqnarray}
\ddot{c}=-\Big(\big(D-1\big)\frac{\dot{a}}{a}+\frac{\dot{c}}{c}\Big)\dot{c}
-\frac{dV_{\rm eff}(c)}{dc}\,,
\end{eqnarray}
where the effective potential is defined as
\begin{eqnarray}
V_{\rm eff}(c)
=\frac{(3+\gamma^2)\big( \Lambda- \frac{2}{1+\gamma^2}\lambda_1\big)}
      {16 \big(\Lambda- \frac{3+\gamma^2}{2(1+\gamma^2)}\lambda_1\big)^2 c^2}
 -\frac{(1-\gamma^2)\Lambda c^2}{4}
+\ln (c)
+{\rm const}\,.
\end{eqnarray}
A typical example of the effective potential is shown in Fig. 1.
There are local minimum at $c=c_2$ and local maximum at $c=c_1$,respectively.
One possible evolution is that from an unstable configuration
to a stable one, namely from the solution of the radius $c_1$ to 
that of $c_2$.
Assuming that correspondingly the brane curvature
changes from $\lambda_1$ to $\lambda_2$ (We now allow for the possibility of
$\lambda <0$),
the flux conservation gives the relation
\begin{eqnarray}
\frac{\lambda_2}{\Lambda}
=\frac{4(1+\gamma^2)^2}{(3+\gamma^2)^2}
\frac{1-\frac{(3+\gamma^2)^2}{4(1+\gamma^2)^2}
\frac{\lambda_1}{\Lambda}}
{1-\frac{2}{1+\gamma^2}\frac{ \lambda_1}{\Lambda}}.
\end{eqnarray}
Noting that $c_i=\big(2\Lambda-2(D-1)\lambda_i/(D-2) \big)^{-1/2}$,
the initial and final radii are related as
\begin{eqnarray}
\frac{c_2^2}{c_1^2}
=\frac{3+\gamma^2}{1-\gamma^2}
\Big(1-\frac{2}{1+\gamma^2}
\frac{\lambda_1}{\Lambda}\Big).
\end{eqnarray}
Clearly for $\lambda_1 >\lambda_{\rm  inst}$, $c_2<c_1$.
Thus, we see that the radius of two-dimensional internal space must become smaller during the evolution, if the unstable configuration evolves to
the stable one.
There is other possibility of cosmological evolution:
the {\it decompactification} $c\to \infty$, namely
to the $(D+2)$-dimensional dS solution with expansion rate given by $(2\Lambda/D(D+1))^{1/2}$. 
These two possibilities may be distinguished in terms of
the initial conditions. If initially $\dot{c}<0$ the radion evolves to the true minimum at $c=c_2$, whereas it may evolve to $c\to \infty$, if initially $\dot{c}>0$. We focus on more realistic the first possibility.

There is also another critical brane expansion rate
\begin{eqnarray}
\frac{\lambda_{\rm AdS}}{\Lambda}
=\frac{4(1+\gamma)^2}{(3+\gamma)^2}.\label{ads}
\end{eqnarray}
Note that always $\lambda_{\rm inst} <\lambda_{\rm AdS}< \lambda_{\rm max}$.
For $\lambda_{\rm inst}< \lambda_1 \leq \lambda_{\rm AdS}$, $\lambda_2\geq 0$
and thus the final $D$-dimensional geometry is dS,
while for $\lambda_{\rm AdS}< \lambda_1 <\lambda_{\rm max}$, $\lambda_2<0$
and the final stable $D$-dimensional geometry is anti- de Sitter (AdS). 
The thermodynamical arguments
cannot be applied to AdS brane configurations
since we should require the reality 
of the {\it temperature} $H^D$ (non-negativity of $\lambda$).
\begin{figure} 
\begin{minipage}[t]{.40\textwidth}
   \begin{center}
    \includegraphics[scale=.35]{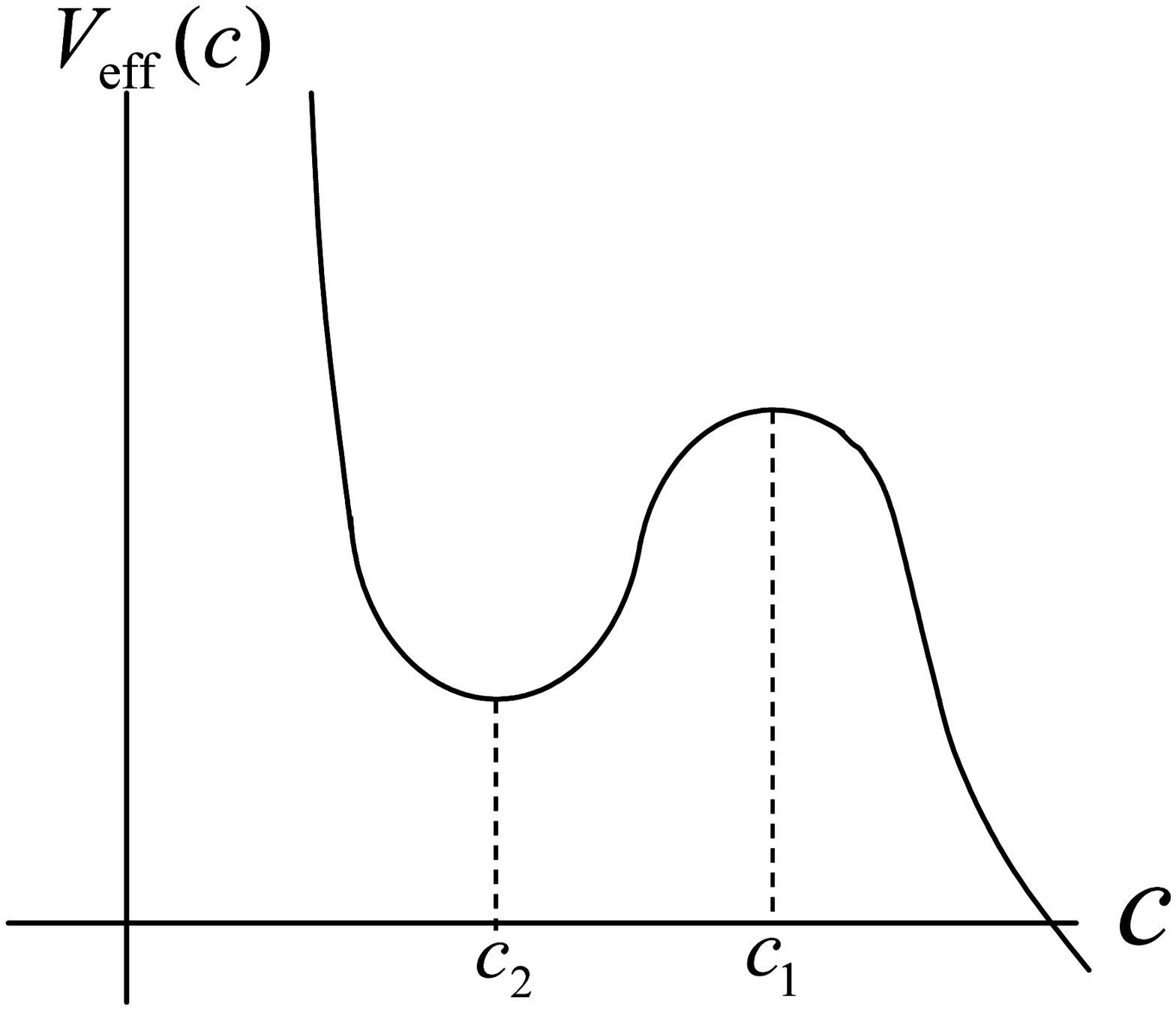}
        \caption{The schematic view of the radion effective potential
is shown.
The local maximum and minimum are located at $c_1$ and $c_2$, respectively.
For $\lambda_1 >\lambda_{\rm inst}$, $c_2 <c_1$ is always satisfied.}
   \end{center}
   \end{minipage}
\hspace{0.1mm}
  \begin{minipage}[t]{.40\textwidth}
   \begin{center}
    \includegraphics[scale=.35]{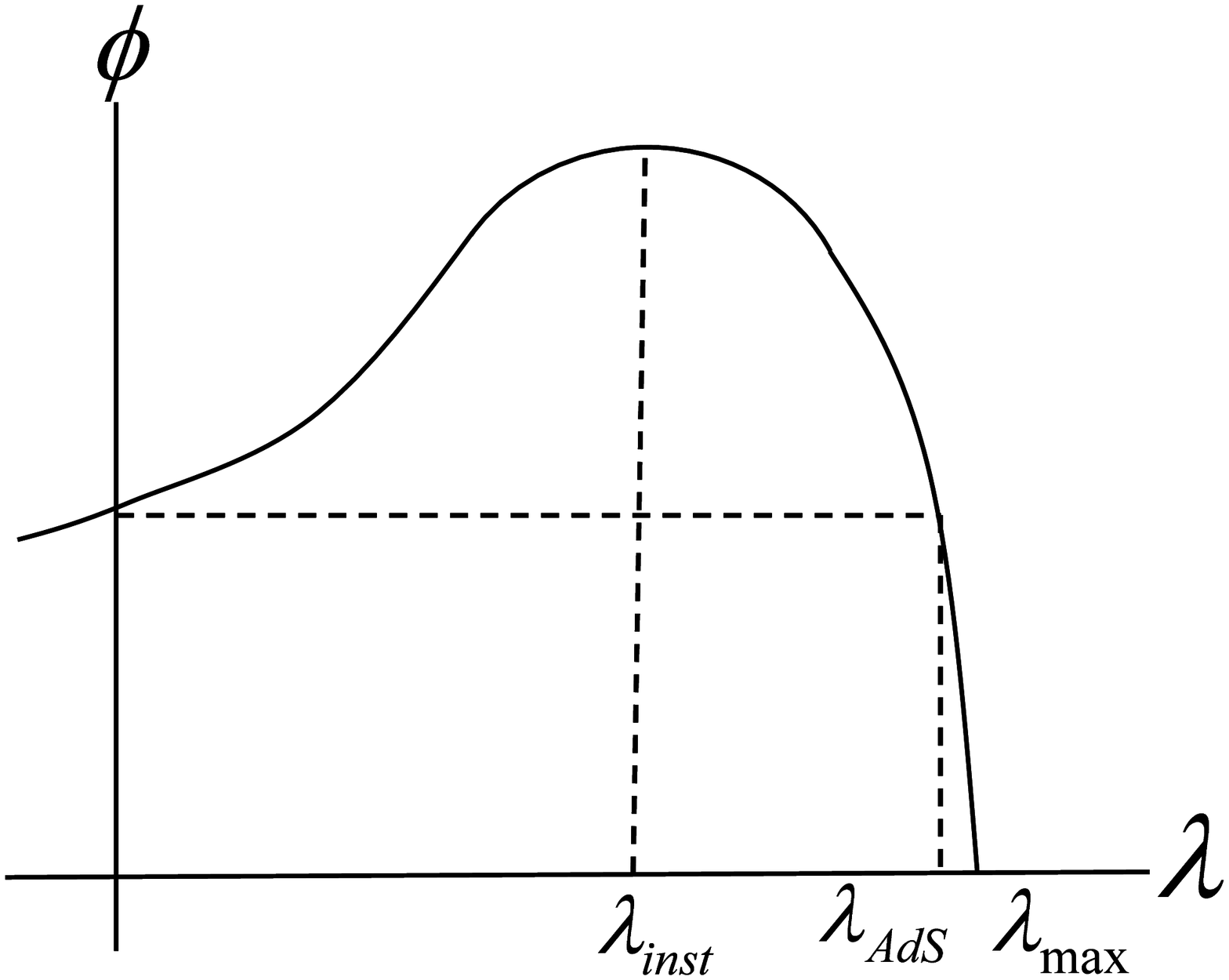}
        \caption{A typical example of the behavior of the magnetic flux Eq.
(\ref{flux2}) is shown as a function $\lambda/\Lambda$.
The sequence can be analytically continued to negative $\lambda$, i.e.
AdS solutions.
Around the maximum value, for the same value of the flux, there are two 
possible values of positive $\lambda$, namely unstable and stable solutions,
implying the evolution between two solutions with conserved flux.}
   \end{center}
   \end{minipage}
\end{figure}
In Fig. 2, we showed a typical example of the magnetic flux as a function
of $\lambda$.
As is seen, for increasing $\lambda$,
the flux takes a maximum value at $\lambda=\lambda_{\rm inst}$, then
starts to decrease and vanishes at $\lambda_{\rm max}$.
For the values of $\lambda_{\rm inst} <\lambda \leq \lambda_{\rm AdS}$,
an unstable solution evolve to another dS configuration.
For the $\lambda_{\rm AdS}<\lambda <\lambda_{\rm max}$, the corresponding
stable configuration is AdS.

The above discussion can be naturally applied to
more general cases $\alpha<1$.
The observer on the (+)-brane cannot change the tension of (-)-brane
and thus, $\phi/\beta_-$, i.e. $\tilde \phi$ defined in the previous
subsection is conserved during the cosmological evolution.
Oppositely for the observer on the (-)-brane and $\phi/\beta_+$
is conserved during the evolution.
In the case $\alpha= 1$, 
the critical expansion rate where dynamical instability appears
is the exactly the point given by $(\partial \tilde \phi/\partial H^D)_{\alpha}
=0$. 
In general, since
\begin{eqnarray}
\Big( \frac{\partial \tilde \phi}{\partial H^D}\Big)_{\eta}
&=&\Big(\frac{\partial \tilde \phi}{\partial H^D}\Big)_{\alpha}
+\Big(\frac{\partial \tilde \phi}{\partial \alpha}\Big)_{H}
\Big(\frac{\partial \alpha}{\partial H^D}\Big)_{\eta}
=\Big(\frac{\partial \tilde \phi}{\partial H^D}\Big)_{\alpha}
-
\Big(\frac{\partial \tilde \phi}{\partial \alpha}\Big)_{H}
\frac{\Big(\frac{\partial \eta}{\partial H^D}\Big)_{\alpha}}
     {\Big(\frac{\partial \eta}{\partial \alpha}\Big)_{H}},
\end{eqnarray}
$(\partial \tilde \phi/\partial H^D)_{\eta}
\neq(\partial \tilde \phi/\partial H^D)_{\alpha}$. 
The special case is that of $\alpha=1$, where
$\big(\partial  \eta/\partial H^D\big)_{\alpha}\to 0$
and hence $(\partial \tilde \phi/\partial H^D)_{\eta}=(\partial \tilde \phi/\partial H^D)_{\alpha}$.
The condition $(\partial \tilde \phi/\partial H^D)_{\alpha}
=0$ gives another curve on the $(\alpha,\lambda)$ plane as 
\begin{eqnarray}
 \frac{\lambda_{\ast}}{\Lambda}
&=&\frac{\alpha^{-2+2/(1+\gamma^2)}(1+\gamma^2)}
{2\big(1-\alpha^{1+2/(1+\gamma^2)}\big)(5-\gamma^2)}
\nonumber\\
&\times&
\Big[
2\alpha^{1+4/(1+\gamma^2)}(1-\gamma^2)
+\big(
-3\alpha^3
+\alpha^{2\gamma^2/(1+\gamma^2)}
+2\alpha^{4+2/(1+\gamma^2)}
 \big)
(3+\gamma^2)
-\alpha^{2+6/(1+\gamma^2)}
(5-\gamma^2)
\Big].
\end{eqnarray}
Note that as special limits
\begin{eqnarray}
 \frac{\lambda_{\ast}}{\Lambda}\Big|_{\gamma\to 0}
=\frac{3+6\alpha+9\alpha^2+6\alpha^3+3\alpha^4+2\alpha^5+\alpha^6}
      {10(1+\alpha+\alpha^2)^2}\,,\quad
\frac{\lambda_{\ast}}{\Lambda}\Big|_{\alpha\to 1}
= \frac{(1+\gamma^2)^2}{3+\gamma^2}\,,
\end{eqnarray}
and hence $\alpha\to 1$
$\lambda_{\rm inst}
=\lambda_{\rm crit}
=\lambda_{\ast}$ as we have seen previously.
It supports that
in the case $\alpha = 1$,
the cosmological evolution keeps the rugby ball shape during the cosmological evolution.
However, in the cases $\alpha <1$, $\lambda_{\rm crit}
\neq \lambda_{\ast}$ and hence it implies that 
the degree of warping $\alpha'$ of the final stable configuration
is different from that of the initial one $\alpha$ in general.

Before closing this section,
in Fig. 3, we showed an example of $\lambda_{\rm max}$, $\lambda_{\rm crit}$
and $\lambda_{\ast}$ on the $(\alpha,\lambda)$ plane.
As we see, in the limit $\alpha\to 1$, $\lambda_{\rm crit}$ and
$\lambda_{\ast}$ coincide while in the opposite limit $\alpha\to 0$,
$\lambda_{\ast}$ agrees with $\lambda_{\rm max}$.
Thermodynamically unstable region corresponds to the region
$\lambda_{\rm crit}<\lambda <\lambda_{\rm max}$ for each $\alpha$.
\begin{figure} 
\begin{minipage}[t]{.60\textwidth}
   \begin{center}
    \includegraphics[scale=.90]{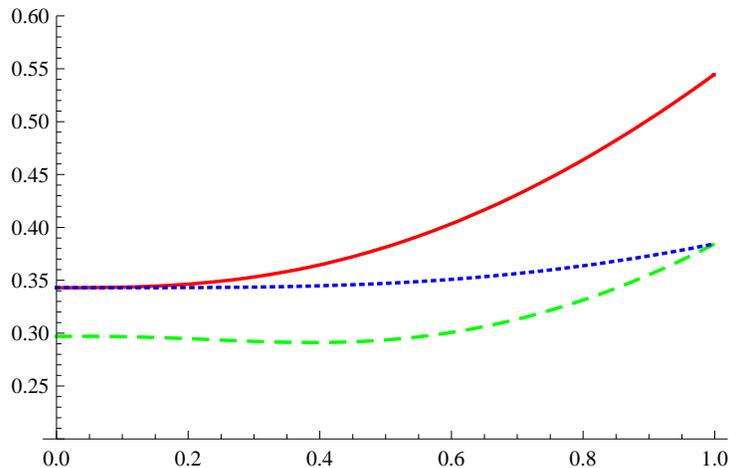}
        \caption{$\lambda_{\rm max}$, $\lambda_{\rm crit}$ and $\lambda_{\ast}$
(divided by $\Lambda$) are shown as a function of $\alpha$ for $\gamma=0.3$,
by solid (red), dashed (green) and dotted (blue) lines, respectively.
The vertical and horizontal axes show $\lambda$ and $\alpha$, respectively.
Note that the solutions in the region $\lambda_{\rm crit}<\lambda<\lambda_{\rm max}$
are thermodynamically (also dynamically) unstable and $\lambda_{\ast}$
is always located in this region.
}
   \end{center}
   \end{minipage}
\hspace{0.1mm}
\end{figure}

\section{Cosmology in six dimensions}

\subsection{Cosmological solutions in six dimensions}

A class of six-dimensional Einstein-Maxwell-dilaton system
has equivalent structure to the $(D+2)$-dimensional Einstein-Maxwell theory
via dimensional reduction.
We rewrite the $(D+2)$-dimensional metric as
\begin{eqnarray}
ds^2_{(D+2)}
=
\underbrace{e^{-(D-4)\phi(x)/2}g_{ab}(x)dx^adx^b}_{6\D}
+\underbrace{e^{2\phi(x)}\delta_{mn}dy^mdy^n}_{(D-4)\D},
\label{metric_form_6+n}
\end{eqnarray}
where $(D-4)$-dimensional part is compactified
and the field strength with
$ \cF_{ab}=\cF_{ab}(x)\quad\text{and}\quad\cF_{mM}=0 $.
With the above ansatz, dimensional reduction on the $(D-4)$
-dimensional manifold yields
\begin{eqnarray}
S^{(6)}&=&\int d^6x\sqrt{-g}
\Big[\frac{1}{2}
\Big(R
-\partial_a\varphi\partial^a\varphi-2e^{-\gamma\varphi}\Lambda\Big)
-\frac{1}{4}e^{\gamma\varphi}F_{ab}F^{ab}
\Big], \label{6d_action}
\end{eqnarray}
with identifications: Eq. (\ref{def_gamma}),
\begin{eqnarray}
\varphi:=\frac{\sqrt{(D-4)D}}{2}\,\phi,
\end{eqnarray} 
and  $F_{ab}:=\cF_{ab}V_{D-4}^{1/2}$
($V_{D-4}$ is the volume of the $(D-4)$-dimensional flat space).
Here $F_{ab}$ denotes the electro-magnetic field strength.
Again we set $M_{6}^4:=M_{D+2}^{D}V_{D-4}=1$, unless it should 
be shown explicitly.
For $\gamma=1$ the action~(\ref{6d_action}) coincides with
the bosonic part of Nishino-Sezgin (Salam-Sezgin) supergravity
(some fields are set to be zero consistently).
From Eq. (\ref{def_gamma}), $0\leq \gamma\leq 1$.
After dimensional reduction, $D$ is a parameter
related to the dilatonic coupling $\gamma$ as Eq. (\ref{def_gamma}),
and hence can be arbitrary real, non-negative number.
In particular, Nishino-Sezgin supergravity
($\gamma=1$) \cite{ns} is reproduced by taking $D\to\infty$.

We also define the 3-brane tension as $\tau_i:=\sigma_i V_{D-4}$.
The tension of the 3-brane does not couple to the dilatonic scalar field after dimensional reduction.
According to the above-mentioned way, 
we derive the cosmological solutions in the six-dimensional
theory Eq. (\ref{6d_action})
from the higher-dimensional dS brane solutions Eq. (\ref{intermsofxi}).
We identify $e^{\phi}=\xi^{(1-\gamma^2)/(1+\gamma^2)}e^{H t}$.
Then the six-dimensional metric is found to be
\begin{eqnarray}
g_{ab}dx^adx^b
=
\xi^{2/(1+\gamma^2)}
\left[-d\tau^2+a^2(\tau)\delta_{ij}dx^idx^j\right]
&+&\frac{b^{2}(\tau)}{2\Lambda}\left[\frac{d\xi^2}{h(\xi)}
+\beta^2h(\xi)d\theta^2\right],
\label{new_6d_sol}
\end{eqnarray}
where the proper time is defined by $\tau=\int e^{(D-4)Ht/4}dt$,
and the scale factors $a(\tau):=e^{D Ht/4}$ and $b(\tau):=e^{(D-4)Ht/4}$,
respectively.
We easily find that
$a(\tau)\propto \tau^{1/\gamma^2}$ and $b(\tau)\propto\tau$,
leading to an accelerating cosmological solution for $\gamma<1$.
The case $\gamma=1$ is exactly the (extensions) scaling solution
in supergravity \cite{6dt,km}.
We also find $\varphi(\tau, \xi)=(2/\gamma)\ln b(\tau)
+(2 \gamma)/(1+\gamma^2)\ln\xi$
and
$F_{\xi\theta}= \big( \beta/\sqrt{2}\big)
\big(Q/\xi^{2(2+\gamma^2)/(1+\gamma^2)} \big)
$, respectively.
Other than this simplest solution
there are two analytically known cosmological solutions 
derived from Kasner-type generalizations of $(D+2)$
dimensional dS brane solutions,
but in the later times both these two solutions approach
the above simplest solution \cite{km}.
Hence the above power-law solution is the late
time attractor.

\subsection{Cosmological implications}

As we mentioned in the previous section,
the class of the Einstein-Maxwell-dilaton theory  Eq. (\ref{6d_action})
has the equivalent structure to the higher-dimensional
Einstein-Maxwell theory via dimensional reduction.
A $(D+2)$-dimensional dS brane solution gives rise to 
the accelerating expanding cosmological solution.
Thus, if a dS brane solution in $(D+2)$ dimensions is unstable,
the corresponding cosmological solution given by Eq. (\ref{new_6d_sol})
is also unstable.
Since we assumed that the $D$-dimensional geometry
keeps the Friedmann-Robertson-Walker form and hence
there are no excitations of any perturbation
mode which depends on the $(D-4)$-dimensional
spatial dimensions, the smooth one-to-one correspondence
of the $(D+2)$-dimensional Einstein-Maxwell theory
with the six-dimensional Einstein-Maxwell-dilaton theory
alway exists at all the times of cosmological evolutions.
If there is a trajectory in the solution space
in the $(D+2)$-dimensional theory,
there also is a corresponding trajectory in the
six-dimensional theory. 
It is also clear that the magnetic flux is conserved 
for the corresponding trajectory in six dimensions.
The mass of an unstable mode scales as
$M_0^2=\propto\mu_0^2/b(\tau)^2\propto \mu_0^2/\tau^2$ from the six dimensional
point of view. 
If the final stable solution in $(D+2)$ dimensional theory is dS brane solution,
then the corresponding cosmological solutions in six-dimensions is also
a accelerating cosmological solution with smaller $\lambda$.
If the final configuration in $(D+2)$-dimensions is AdS,
then the corresponding six-dimensional solution would be a collapsing Universe,
though there may be no analytic form
since there is analytic description of AdS spacetime with the form
of a flat Friedmann-Robertson-Walker metric.

The meaning of $\lambda$ for a four-dimensional observer becomes clearer 
from the perspectives of the four-dimensional effective theory.
From the original $(D+2)$-dimensional metric
\begin{eqnarray}
ds^2_{(D+2)}
=
\xi^{2(1-\gamma^2)/(1+\gamma^2)}
\Big(
\underbrace{e^{-(D-4)B(x^{\mu})/2}q_{\mu\nu}(x)dx^{\mu}dx^{\nu}}_{4\D}
+\underbrace{e^{2B(x^{\mu})}\delta_{mn}dy^mdy^n}_{(D-4)\D}
\Big)
+\frac{\xi^{-2\gamma^2/(1+\gamma^2)}}{2\Lambda}
\Big[
\frac{d\xi^2}{h(\xi)}+\hat\beta^2h(\xi)d\theta^2
\Big],
\end{eqnarray}
and magnetic field given in Eq. (\ref{fs}),
the four-dimensional effective theory, composed of the 
four-dimensional metric $q_{\mu\nu}(x^{\mu})$ and the modulus $B(x^{\mu})$,
for the observer on the $(+)$-brane is obtained.
Then, after a conformal transformation $\hat q_{\mu\nu}= e^{2\gamma^2 B/(1-\gamma^2)} q_{\mu\nu}$,
we can go to the Einstein frame.
By defining the canonically normalized modulus
$\chi:= (2\gamma\sqrt{2(1+\gamma^2)}/(1-\gamma^2))B$,
we obtain the four-dimensional effective Einstein-scalar theory with an exponential
potential 
\begin{eqnarray}
S_{\rm eff}
&=&\int d^4 x
\sqrt{-\hat q}
\Big[
 \stackrel{(4)}{\hat R}
-\frac{1}{2}
 {\hat q}^{\mu\nu}\partial_{\mu}\chi
           \partial_{\nu}\chi
-2\lambda e^{-\sqrt{2\gamma^2/(1+\gamma^2)}\chi} 
\Big]\,.\label{eff}
\end{eqnarray}
Thus, $\lambda$ characterizes the potential of an effective
quintessential scalar field.
The cosmic scale factor in the Einstein frame $\hat a\propto
a^{1+\gamma^2}$ is proportional to $\hat \tau^{(1+\gamma^2)/(2\gamma^2)}$,
where $\hat \tau$ is the cosmic proper time in the Einstein frame.
This scalar field may be the dark energy source, which accelerates
the current Universe.
A cosmological evolution in higher dimensions leads the geometry
to stable solution with a smaller $\lambda$.
The cosmological evolution in higher dimensions may be seen
as a process of the transition from the initial cosmological inflation to the 
current dark energy dominated Universe from the four-dimensional
perspectives.

\section{summary}

We discussed the stability of a de Sitter (dS)
brane solutions in the higher-dimensional Einstein-Maxwell theory. 
We confirmed that the dS brane solutions with relatively high
expansion rates are unstable against scalar-type linear perturbations
with respect to the symmetry of dS spacetime.
Such an instability commonly appears in the wide class of compactifications
with an external dS spacetime and can be understood as a type of radionic instabilities.

Then, we generalized a
relation found in the six-dimensional dS brane solutions Ref. \cite{ksm}, 
which has very similar structure to the ordinary laws of thermodynamics,
to the case of the higher dimensions as Eq. (\ref{1st}).
In this relation, the area of dS horizon (integrated over two internal dimensions)
behaves the thermodynamical entropy.
The area of dS horizon is essentially double-valued function of the
flux and (one of) the brane tensions. 
A dynamically unstable solution is also in the thermodynamically unstable branch.
The boundary between the thermodynamically stable and unstable branches is given
by a curve, where one-to-one map
from the plane of model parameters $(\alpha,\lambda)$,
where $\alpha$ controls the degree of warp and $\lambda$
characterizes the curvature of dS spacetime,
to the conserved quantities breaks down.
These are the generalizations of the results of the 
case of six dimensions discussed in \cite{ksm} to higher dimensions.

Then, we discussed the possible cosmological evolutions.
We firstly focused on the case $\alpha=1$, the case of an exactly rugby ball shaped
bulk.
In this case, during the cosmological evolution,
the bulk keeps the rugby-ball type shape.
There are two possibilities of the evolutions of 
unstable cosmological solutions:
One possibility is to settle down a stable configuration.
The other one is that the compactified two
extra-dimensions are decompactified.
This strongly depends on the initial condition
and if the internal space is initially shrinking,
the final configuration is another stable dS brane solution.
Then, the size of the internal space in the final
configuration usually is smaller than the initial 
unstable configuration. 
Furthermore, if initially $\lambda_{\rm inst}<\lambda<\lambda_{\rm AdS}$,
where $\lambda_{\rm inst}$ and $\lambda_{\rm AdS}$ are given
in Eq. (\ref{yoko}) and Eq. (\ref{ads}), the final configuration
is also a dS brane solution, while $\lambda_{\rm AdS}<\lambda<\lambda_{\rm max}$,
the final configuration is an AdS brane solution.
This picture can be easily generalized to the case
of unstable dS configurations with non-trivial magnetic flux.
In this case of the warped bulk $\alpha<1$, the basic behavior remains the same as
the rugby ball case:
the initially unstable dS brane solution 
evolves toward the stable dS/AdS brane solutions
(or decompactified).
For the observer on the (+)-brane, during the
cosmological evolution $\phi/\beta_-$ is conserved
while for the observer on (-)-brane,
$\phi/\beta_+$ is conserved.
In general, the shape of the bulk of the final stable
configuration is different from 
that of the initial stable one.

Finally, we have discussed the stability of brane
cosmological solutions
in the six-dimensional Einstein-Maxwell-dilaton theory Eq. (\ref{6d_action}),
including the Nishino-Sezgin (Salam-Sezgin), gauged supergravity as the special limit.
This class of six-dimensional Einstein-Maxwell-dilaton theory 
has equivalent structure to the $(D+2)$-dimensional
Einstein-Maxwell theory and the number of dimensions $D$
affects the dilatonic coupling in the reduced six-dimensional theory.
Thus, if the seed dS brane solution is unstable in $(D+2)$-dimensional spacetime,
the corresponding cosmological solution in six dimensions is also unstable:
if the final configuration is another dS brane solution in $(D+2)$ dimensions,
this also may be true in six dimensions.
The cosmological evolution in higher dimensions may be seen
as a process of the transition from the initial cosmological inflation to the 
current dark energy dominated Universe from the four-dimensional
perspectives.

\section*{Acknowledgement}
The author thanks Prof. D. Langlois for comments.
The author also wishes to thank the anonymous reviewers for his/her comments.
This work was partially supported by the project TransRegio 33 {\it The Dark Universe}.

\appendix

\end{document}